\documentclass[draft]{mn2e}
\usepackage{graphicx}
\usepackage{epstopdf}
\usepackage{subfigure}
\usepackage{lscape}
\usepackage{longtable}
\usepackage{wasysym}
\usepackage{psfig}

\usepackage{wasysym}

\def\gs{\mathrel{\raise0.35ex\hbox{$\scriptstyle >$}\kern-0.6em
\lower0.40ex\hbox{{$\scriptstyle \sim$}}}}
\def\ls{\mathrel{\raise0.35ex\hbox{$\scriptstyle <$}\kern-0.6em
\lower0.40ex\hbox{{$\scriptstyle \sim$}}}}
\def\ls{\mathrel{\hbox{\rlap{\hbox{\lower4pt\hbox{$\sim$}}}\hbox{$<$}}}}
\def\gs{\mathrel{\hbox{\rlap{\hbox{\lower4pt\hbox{$\sim$}}}\hbox{$>$}}}}

\def\mnras {{\sc MNRAS}}

\title[Backsplash Galaxies in Isolated Clusters]
      {Backsplash Galaxies in Isolated Clusters}
\author[K.\,A.\,Pimbblet ]
       {Kevin A.\ Pimbblet \thanks{Email: Kevin.Pimbblet@monash.edu}
        \vspace*{1mm}\\
School of Physics, Monash University, Clayton, Victoria 3800, Australia}

\date{\fbox{\sc Draft: \today\ --- Do Not Distribute}}

\pagerange{000--000}

\begin{document}

\maketitle

\begin{abstract}
At modest radii from the centre of galaxy clusters, individual
galaxies may be infalling to the cluster for the first time,
or have already visited the cluster core and are coming back
out again.  This latter population of galaxies is known as the
backsplash population. Differentiating them from the infalling
population presents an interesting challenge for observational 
studies of galaxy evolution.  To attempt to do this, we assemble
a sample of 14 redshift- and spatially-isolated galaxy clusters
from the Sloan Digital Sky Survey.
We clean this sample of cluster-cluster mergers to ensure
that the galaxies contained within them are (to an approximation)
only backsplashing from the centre of their parent clusters and 
are not being processed in sub-clumps. By stacking them 
together to form a composite cluster, we find evidence for
both categories of galaxies at intermediate radii from the 
cluster centre.
Application of mixture modelling to this sample then
serves to differentiate the infalling galaxies 
(which we model on galaxies from the cluster outskirts)
from the backsplash ones 
(which we model on galaxies in the high density core with
low velocity offsets from the cluster mean).
We find that the fraction of galaxies with populations
similar to the low velocity cluster core galaxies is
$f = -0.052R/R_{virial} + 0.612 \pm 0.06$
which we interpret as being the
backsplash population fraction at $1<R/R_{virial}<2$.  
Although some interlopers may be affecting our results, 
the results are demonstrated to be in concordance with
earlier studies in this area that support 
density-related mechanisms as being the 
prime factor in determining the 
star formation rate of a galaxy.
\end{abstract}


\section{Introduction}
During the course of its lifetime, 
a galaxy may travel through several different environments
and be affected by different physical mechanisms in each of them.
As a very simplistic, illustrative example, 
consider an `average' test galaxy that gets
accreted on to a filament of galaxies.
As the galaxy flows along the filament, it may undergo some
pre-processing due to first time harassment 
(Moore et al.\ 1996) caused by the increase
in local galaxy density (Porter et al.\ 2008).
The filament will eventually funnel the galaxy to anisotropically
fuel the growth of a cluster of galaxies (Pimbblet 2005; see also White,
Cohn \& Smit 2010).
If the star formation rate of the galaxy has not already
undergone significant transformation due to (slow) gas starvation
(cf.\ van der Wel et al.\ 2010;
Balogh, Navarro, \& Morris 2000; 
Larson, Tinsley, \& Caldwell 1980), then the cluster environment
will almost certainly act to truncate it to much lower levels (e.g.\ 
Lewis et al.\ 2002; G{\'o}mez et al.\ 2003; Kauffman et al.\ 2004;
Pimbblet et al.\ 2006) as higher pressure and
higher density mechanisms such as ram pressure stripping 
(Gunn \& Gott 1972) and tidal interaction with the cluster 
gravitational potential (Byrd \& Valtonen 1990; Rose et al.\ 2001)
become more important. Additionally, ram pressure stripping may
also induce a final star star formation episode in a galaxy 
to remove any remaining gas (Smith et al.\ 2010a).  

These mechanisms may also change the morphology 
of the galaxy by (e.g.) perturbation of spirial arms 
(Quilis, Moore, \& Bower 2000) and the colour will
get progressively redder as the 
star-formation rate declines and the
bluer stellar population dies off.
Such processes are generally posited to contribute to 
the well-known morphology-density and 
density-star formation rate relationships
(e.g.\ Dressler 1980; Smith et al.\ 2005;
Holden et al.\ 2007).  
Taken together, the origin of 
these relations appear dependant on 
a variety of physical 
mechanisms that act on distinct galaxy 
masses, at different epochs, and over certain time-scales.

Attempting to analyze the star formation rates of such 
galaxies in order to extract more general conclusions 
about the driving mechanisms behind galaxy evolution 
and the morphology-density relation is not 
without its trials.  For example, the work of Porter et al.\ (2008)
examines the star-formation rate of galaxies as a function 
of cluster-centric radius along the vector of a filament going
away from a cluster core.  They suggest there is a peak in 
star-formation rate at a few Mpc from the cluster core along
filaments driven mostly by light from dwarf galaxies -- 
this is something that would not be seen 
if one simply averages all vectors away from the cluster 
core, as works such as 
Lewis et al.\ (2002), G{\'o}mez et al.\ 2003, 
and Pimbblet et al.\ (2006) do.

One further complication to addressing the changes in star 
formation rate (whether radially averaged, or chosen along the
vector of a filament as in Porter et al.\ 2008) is the 
presence of \emph{backsplash} galaxies.
Backsplash galaxies are those galaxies that have 
already entered the core regions of a galaxy cluster and
have subsequently come back out again 
(Gill, Knebe, \& Gibson 2005; 
see also Ludlow et al.\ 2009; Warnick et al.\ 2008; 
Aubert \& Pichon 2007).  
The existence of backsplash
galaxies is not unexpected: Balogh et al.\ (2000)
find that 50$\pm$20\% of particles in the region of 
$R_{200}$ to $2R_{200}$ from a cluster have passed within
$R_{200}$.  Mamon et al.\ (2004) further compute that the
maximum distance that a backsplash particle has 
is $2.5R_{100}$, and for a galaxy is $1.7R_{100}$.
This is supported by Moore et al.\ (2004) who compute that
half of all halos between $R_{virial}$ and $2R_{virial}$
have passed through their parent cluster, many directly 
through the high density core regions of clusters
Interestingly, Knebe et al.\ (2008) show that the backsplash
population is predicted to exist in more than one 
cosmology.

The implication of these findings is observationally 
concerning: to understand the origin of 
(e.g.) the morphology-density
relation and the mechanisms that drive galaxy evolution, 
we should in principle
objectively disentangle the backsplash galaxy population 
from those that are infalling to the cluster for the first time.

Gill et al.\ (2005) provide a succinct observational
method to detect the presence of a backsplash
population in galaxy clusters.  Given that
the backsplash population will have a distinct
velocity distribution, observers need simply to
examine the distribution in a suitably large
radius interval from the cluster centre, say
between 1 and 2 virial radii.  A backsplash population
will exhibit a centrally peaked distribution for
this interval, whereas a pure infalling population
will possess a higher modal value.

A number of observational studies have
now confirmed the
existence of a backsplash population in
galaxy clusters 
(e.g.\ Aguerri \& Sanchez-Janssen 2010;
Smith et al.\ 2010b;
Sato \& Martin 2006;
Pimbblet et al.\ 2006;
Rines et al.\ 2005;
Sanchis et al.\ 2004). Yet,
although these studies strongly suggest that
the backsplash population exists,
they have not in detail
disentangled the backsplash population
from the infalling one, although using 
galaxies with different spectral features
appears to hold much promise 
for this (see Pimbblet et al.\ 2006;
Rines et al.\ 2005).

This work revisits the backsplash concept 
and (observationally) attacks the question
of how the backsplash fraction is expected to vary 
with radius from the cluster centre.
In Section~2, we construct a sample of isolated galaxy
clusters from the Sloan Digital Sky Survey to work 
with.  This sample should be free of significant nearby
structures and sub-clumps that may be acting to 
pre-process before they reach the centre of their parent
galaxy clusters.
In Section~3 we observe the presence of the backsplash 
population in our sample.  We apply a mixture model
to deduce how the fraction of backsplash galaxies varies
with radius from the centre of clusters and compare
our result with previous studies in this area.
In Section~4 we generalize our results and discuss
a number of caveats about our work before summarizing
the main findings in Section~5.

Throughout this work, we adopt 
H$_0 = 100$ $h_{100}$\,km\,s$^{-1}$\,Mpc$^{-1}$,
$\Omega_M=0.27$, and $\Omega_{vacuum}=0.73$.

\section{Sample}
For this work, we wish to use a sample of galaxy
clusters that are free from significant subclustering
that may skew some of the earlier
observational results.  For instance, Pimbblet
et al.\ (2006) make use of very X-ray bright, massive
clusters, some of which exhibit significant
substructure (e.g.\ A1664; A3888).  
The issue is that inside these
sub-clusters, galaxies may have been already
effected by various evolutionary mechanisms as
if they had fallen into a cluster already.  We wish
to avoid this scenario as much as possible.
Fortuitously, such an ensemble already exists:
Plionis et al.\ (2009) have assembled a large
sample of galaxy clusters from Abell, 
Corwin \& Olowin (1989) that are `clean' 
of merging and interacting clusters.  This
sample should be almost ideal for 
studying a backsplash population
with, since in principle 
it should be free from coherent infall
(e.g.\ large groups and sub-clusters of galaxies) and 
relatively isolated in space.

From the sample of Plionis et al.\ (2009) we
select only those clusters that are contained within
Sloan Digital Sky Survey (SDSS; Abazajian et al.\ 2009)
boundaries and appear in the SDSS C4 catalogue
(Miller et al.\ 2005) to study further -- this is 
a matter of convenience given SDSS has good photometry
down to a reasonably faint limit for nearby 
galaxy clusters, coupled with quality spectra.
Moreover, by using SDSS as our exclusive 
data source, we try to actively 
avoid any biases that might be present in 
ensembles from inhomogeneous data sources.
These clusters are detailed in Table~\ref{tab:sample}.
Given that all of them have been detected by the
C4 cluster finding algorithm 
(Miller et al.\ 2005), they are not chance
superpositions along the line of sight as a 
non-negligible fraction 
of Abell clusters have been shown to be (cf.\ 
Pimbblet et al.\ 2010;
Frenk et al.\ 1990; Lucey 1983).

From this sample, we discard A1452.  The reason
for doing so is that in the C4 catalogue of
Miller et al.\ (2005), this cluster is noted
to have a high probability of
having sub-clustering, as determined from
a Dressler \& Shectman (1988) test, despite 
its inclusion in the Plionis et al.\ (2009)
sample.
Additionally, A0095
is also flagged as not being highly isolated in
redshift space (based on Miller et al.\ 2005's 
`structure contamination flag').  We therefore
discard this cluster as well.
This leaves us with a sample of 14 galaxy clusters
to work with.

We detail the global parameters of these clusters
in Table~\ref{tab:sample} (e.g.\ mean redshift, 
$\overline{cz}$; velocity dispersion, $\sigma_{cz}$; 
X-Ray luminosity in the 0.1--2.4 keV band, $L_X$).
We note that the values presented in Table~\ref{tab:sample}
(i.e.\ sourced from Miller et al.\ 2005)  
compare very favourably to other, sometimes more
extensive and deeper, studies of galaxy clusters 
(e.g.\ A1650 also appears in the Pimbblet 
et al.\ 2006 sample and exhibits completely
consistent values for $\sigma_{cz}$ as used here).
From the values of $\sigma_{cz}$, we compute the
virial radius ($R_{virial}$) for each cluster
from the expression given by Girardi et al.\ (1998):
$R_{200} \sim R_{virial}=0.002\sigma_{cz}$ (where
$R_{200}$ is the radius at which the mean interior 
density is 200 times the critical density and
is well approximated by $R_{virial}$).  
Although other
expressions exist for similar radii 
(e.g.\ Carlberg et al.\ 1997 give $R_{virial} \sim 
R_{200} = 3^{1/2} \sigma_{cz} / 10 H(z)$, assuming
an isothermal sphere), this 
expression will suffice for ultimately placing 
our clusters on to a common metric.

We note that the 
sample spans a factor of 1000 in X-Ray luminosity;
from clusters that are often thought of as being 
amongst some of the most massive in the local 
Universe (A1650; cf.\ Pimbblet et al.\ 2006) to
those whose luminosity is comparable to rich
groups, or lower.  
The velocity dispersion of these clusters,
however, is more in-line with that of intermediate 
to massive clusters 
(see Table~2 of Pimbblet et al.\ 2006) with a 
median value of 849 kms$^{-1}$ (see also
Ortiz-Gil et al.\ 2004 who discuss the 
peculiarities of the $L_X$--$\sigma_{cz}$
relationship in depth).  
This implies that the range in total mass
of our clusters is likely to be little more
than a factor of $\sim2$ 
assuming $\sigma_{cz}$ is a reasonable proxy 
for mass.
The range of
$\sigma_{cz}$ spans a factor of $\sim$2.4.  This
is shortened to only a factor of $\sim1.6$ when
the highest and lowest values in the sample are 
ignored.  
In terms of look-back time, the clusters span 
$\approx$1 Gyr range at maximum 
(from A0119 to A0655), or
an inter-quartile range of $\approx$0.3 Gyr -- 
short enough that the time-scales for 
environmental processes driving galaxy evolution
will not greatly affect the sample 
(e.g.\ Kodama \& Smail 2001).  Or put another
way: the clusters should all be at a very similar 
stage in their evolution.

%
%
\begin{table*}
\begin{center}
\caption{The sample of Abell clusters used in this work.  The mean redshift ($\overline{cz}$),
and velocity dispersion ($\sigma_{cz}$) are sourced from Miller et al.\ (2005).  
The coordinates specify the Vizier position of the cluster.
The X-Ray luminosities 
($L_X$) are sourced, if available, from the Base de Donn\'{e}es Amas de Galaxies X
(BAX; webast.ast.obs-mip.fr/bax; Sadat et al.\ 2004) which in turn derives its data from
other literature sources (Popesso et al.\ 2007; Reiprich \& B\"{o}hringer 2002;
see also Ebeling et al.\ 1996).  The virial radius ($R_{virial}$) is computed
from $\sigma_{cz}$; see text for details.
\hfil}
\begin{tabular}{lllllll}
\noalign{\medskip}
\hline
Name  & RA      & Dec     & $L_X$                           & $\overline{cz}$ & $\sigma_{cz}$  & $R_{virial}$  \\
      & (J2000) & (J2000) & ($\times 10^{44}$ erg s$^{-1}$) & (km\,s$^{-1}$)  & (km\,s$^{-1}$) & (Mpc)      \\
\hline
A0023           & 00 21 11.7 & $-$00 57 40 & N/A   & 31854 & 558  & 1.12 \\
A0095$\dagger$  & 00 45 56.8 & $-$00 54 30 & N/A   & 32736 & 611  & 1.22 \\
A0119           & 00 56 14.3 & $-$01 08 40 & 3.296 & 13344 & 785  & 1.57 \\
A0655           & 08 25 17.6 & $+$47 10 09 & 6.651 & 38136 & 813  & 1.63 \\
A1205           & 11 13 58.1 & $+$02 29 56 & 1.774 & 22506 & 938  & 1.88 \\ 
A1346           & 11 41 11.8 & $+$05 44 05 & 0.371 & 29472 & 852  & 1.70 \\
A1364           & 11 43 27.1 & $-$01 43 51 & 0.071 & 31833 & 599  & 1.20 \\
A1424           & 11 57 26.4 & $+$05 05 52 & 0.866 & 22764 & 780  & 1.56 \\
A1452$\ddagger$ & 12 03 19.6 & $+$51 42 16 & N/A   & 18471 & 762  & 1.52 \\
A1496           & 12 13 07.8 & $+$59 16 03 & 0.059 & 28803 & 426  & 0.85 \\
A1620           & 12 50 03.0 & $-$01 33 45 & 0.004 & 25275 & 1007 & 2.01 \\
A1650           & 12 58 34.7 & $-$01 43 15 & 6.991 & 25176 & 864  & 1.73 \\
A1767           & 13 36 31.6 & $+$59 08 51 & 2.429 & 21111 & 988  & 1.98 \\
A2026           & 15 08 32.0 & $-$00 16 03 & 0.253 & 27186 & 846  & 1.69 \\
A2048           & 15 15 20.0 & $+$04 23 05 & N/A   & 29451 & 971  & 1.94 \\
A2670           & 23 54 13.7 & $-$10 25 09 & 2.281 & 22836 & 976  & 1.95 \\
\hline
\multispan{7}{$^\dagger$A0095 is removed from our sample on the basis of a high structure contamination} \\
\multispan{2}{flag (Miller et al.\ 2005).} \\
\multispan{7}{$^\ddagger$A1452 is removed from our sample since Miller et al.\ (2005) indicate a high probability} \\
\multispan{2}{of subclustering.}  \\
\noalign{\smallskip}
\end{tabular}
  \label{tab:sample}
\end{center}
\end{table*}

Consequentially, we
suggest that the sample contains
reasonably self-similar clusters that should have
comparable backsplash populations for the purposes
of this study.

\section{Backsplash in the Composite Cluster}

%
%
\begin{figure}
\centerline{\psfig{file=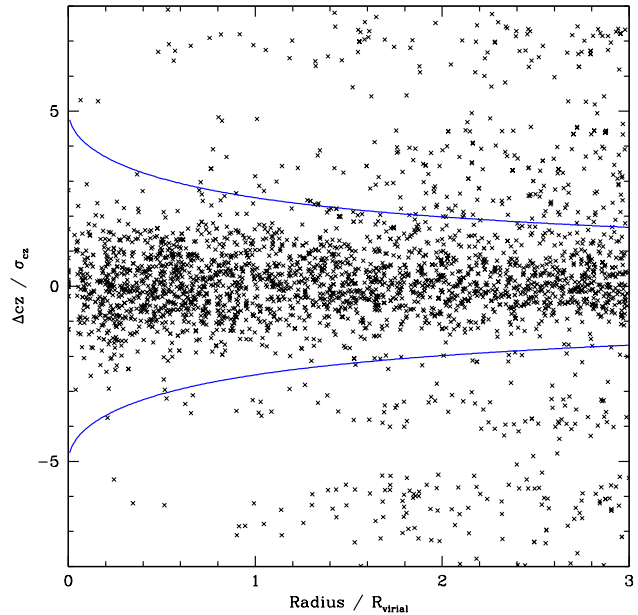,angle=0,width=3.5in}}
  \caption{The composite, stacked cluster plotted as the
difference in velocity normalized to each cluster's individual
velocity dispersion as a function of virial radius.
The solid curves denote the 3$\sigma$ contour (caustics) of the
mass model of Carlberg, Yee \& Ellingson (1997).  Only galaxies
inside these caustics are considered members and
used in the subsequent analysis.
}
\label{fig:stacked}
\end{figure}

%
%
\begin{figure}
\centerline{\psfig{file=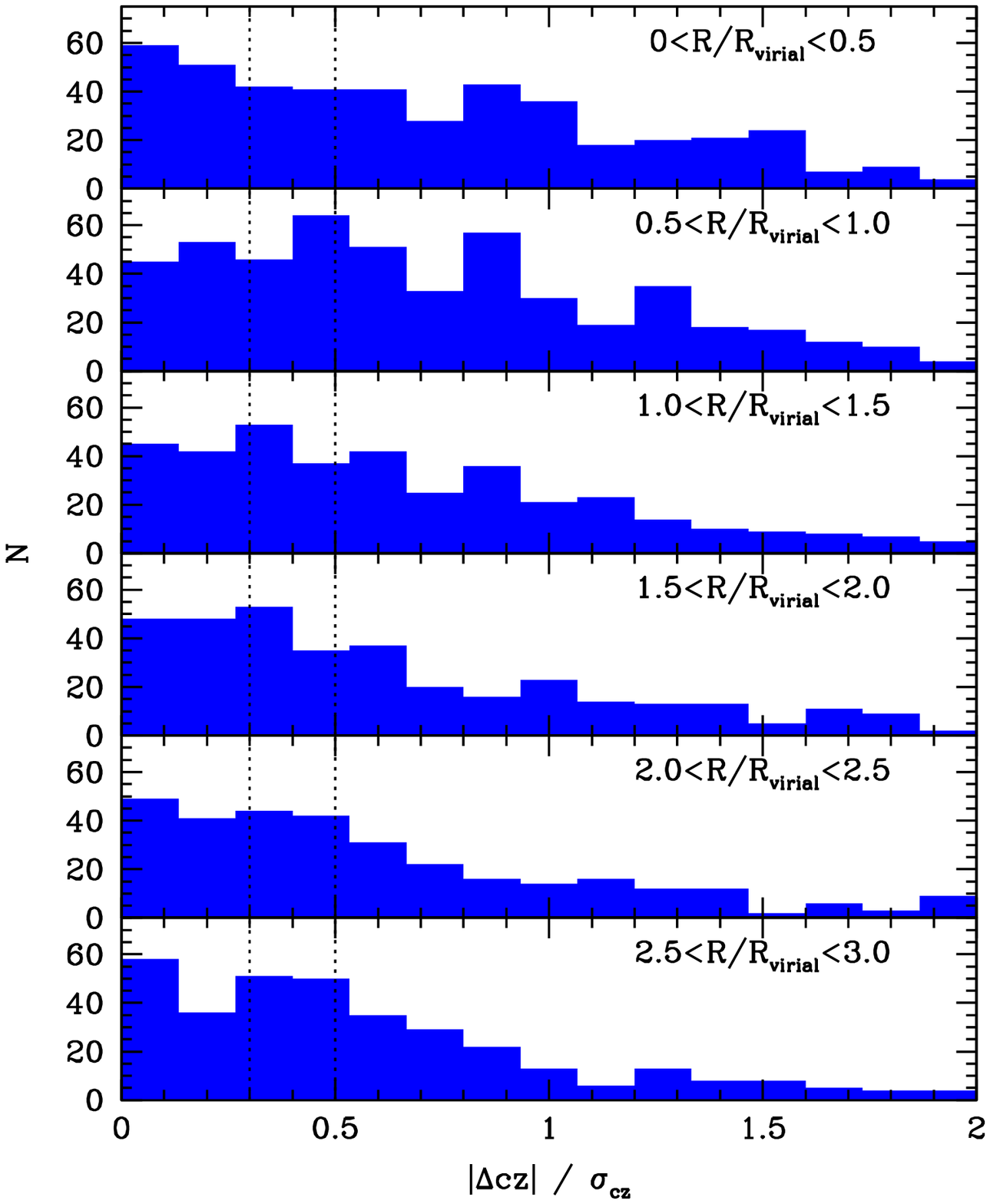,angle=0,width=4.6in}}
  \caption{Histograms of $|\Delta cz| / \sigma_{cz}$ in the
composite cluster, split in to radial bins.
The mode of the distribution is expected to reside between
the vertical dotted lines for a pure population of first
time infallers.  A mixture of both
backsplash and infalling populations can be seen
at low to intermediate radii.
}
\label{fig:histogs}
\end{figure}

To proceed, we now extract all galaxies from SDSS that are
within $3R_{virial}$ of their cluster centres, as 
given in Table~\ref{tab:sample}.  
The homogeneity of the sample is then exploited in order
to construct a stacked, composite cluster.  This is
done by placing the cluster on to a standard fixed metric
through the use of the computed $R_{virial}$ values
(Table~\ref{tab:sample}) and scaling each cluster 
by its velocity dispersion.  The result of this stacking
process is illustrated in Fig.~\ref{fig:stacked} where
all galaxies within $3R_{virial}$ are plotted against
their scaled velocity difference from $\overline{cz}$.
To better define the extent of the cluster, we follow
Pimbblet et al.\ (2006; see also Lerchster et al.\ 2010) 
and utilize the cluster mass model of Carlberg, Yee \&
Ellingson (1997; see also Carlberg et al.\ 1997; 
Rines \& Diaferio 2010 and references therein).
Only galaxies inside the mass model's caustics (Fig.~\ref{fig:stacked})
are considered members and used in our subsequent analysis.
The relative isolation of the composite cluster 
in both a radial and recession 
velocity sense can clearly 
be seen from Fig~\ref{fig:stacked}, even out
to modestly large radii from the cluster centre;
thereby validating our approach of 
using a reduced Plionis et al.\ (2009) sample.

Next, we split the sample in to a number of radial bins. 
Following Gill et al.\ (2005), we create histograms
of $|\Delta cz| / \sigma_{cz}$ in each of these
radial bins and display these
distributions in Fig.~\ref{fig:histogs}. A pure
backsplash population should appear centrally peaked
in these plots, as per the predictions of Gill et al.\ (2005),
whilst an infalling population will be peaked in
the $0.3<|\Delta cz| / \sigma_{cz}<0.5$ region (which we
note is contained inside the mass model limits at all
radii explored; Fig.~\ref{fig:stacked}).  
Indeed,
the infalling population can be expected to have approximately a factor
of two greater relative velocity than the backsplash ones 
(Gill et al.\ 2005) which makes the two populations kinematically
distinct.  Gill et al.\ (2005) report that if all galaxies at modest
radii ($\sim$ few $R_{virial}$) are infalling, then their absolute 
velocity has a modal value of $|cz - \overline{cz}|\approx400$ kms$^{-1}$. 
For clusters such as those used in this work (and also 
Pimbblet et al.\ 2006 \& Rines et al.\ 2005) this corresponds to
approximately $0.3<|\Delta cz| / \sigma_{cz} <0.5$.
Fig.~\ref{fig:histogs} shows evidence for both types
of populations being present, similar to previous observational
studies (Rines et al.\ 2005; Pimbblet et al.\ 2006).
At the centre of
the composite cluster ($R/R_{virial}<0.5$), 
the distribution is markedly centrally concentrated in 
$|\Delta cz| / \sigma_{cz}$ terms.  The
apparent degree of central concentration decreases with
increasing radius.  In the zone where we expect a 
mixed population (e.g., $1.0<R/R_{virial}<1.5$), we see that 
the mode of the distribution sits in the range
$0.3<|\Delta cz| / \sigma_{cz}<0.5$, as would be expected 
from an infalling population (Gill et al.\ 2005), along with
many galaxies under this range (i.e.\ the backsplash population).  
As we go out to higher radii, the distinction of the modal 
value lessens.  Furthermore, at large radii, we do not expect
any backsplash galaxies to be present (Mamon et al.\ 2004).

\subsection{Specific Star Formation Rates}

In addition to the difference in the dynamics of
the infall and backsplash population, a difference
in star formation histories may also be 
present.  If the backsplash population has
already experienced the hostile high density
core regions of a galaxy clusters (cf.\ Moore et al.\ 2004), 
then it follows that their star formation rate
distribution should reflect this.  This provides us with a
method to potentially deduce the minimum fraction of 
backsplash galaxies present in the
mixed population (Fig.~\ref{fig:histogs}).
The reason that it will be a minimum fraction is
that not 100\% of the backsplash galaxies will have
experienced star-formation truncation comparable to
core-region galaxies -- some may have more fortunate
orbits around the cluster centre and retain an appreciable
star formation rate. Indeed, this is supported by 
Rines et al.\ (2005) who demonstrate that some backsplash
galaxies must be emission line galaxies.  Hence 
the status of a given galaxy as being a backsplash one may
not be the prime determinant of their star formation.
Therefore in the analysis below we examine the 
specific star formation rate \emph{distribution}
of a large number of galaxies at various locations inside
the Carlberg, Yee \& Ellingson (1997; Fig.~\ref{fig:stacked})
cluster membership caustic.

If the galaxy population located at 
$0.3<|\Delta cz| / \sigma_{cz}<0.5$ is considered to be
a mixture of both infalling and backsplash galaxies,
then we should be able to model it as a superposition
of both and determine the how the fraction of backsplash
galaxies varies as a function of radius from the cluster centre
(cf.\ Rines et al.\ 2005; see also Pimbblet et al.\ 2006).

Arguably the simplest population to get a handle on 
is the infalling population.
The simulations of Mamon et al.\ (2004) demonstrate almost 
no backsplash galaxy will reside at $>1.7 r_{100}$.
We therefore assume that at radii greater than 
$2 R_{virial}$ (Fig.~\ref{fig:histogs}), 
the galaxy population consists of 
approximately pure first-time infallers.
For galaxies in the range 2--3 $R_{virial}$, 
we find the logarithm of the mean specific star formation rate
(SSFR; in units of $M_{\astrosun} yr^{-1} M^{\star -1}$; see 
Brinchmann et al.\ 2004 and
www.mpa-garching.mpg.de/SDSS for details of this parameter's derivation)
is $-10.99 \pm 0.88$ with an inter-quartile range of 1.79. 
If we restrict this
sample to $0.5<|\Delta cz| / \sigma_{cz}<2.0$ (i.e.\ well outside
the centrally concentrated peak one might expect for
backsplash galaxies, and outside the plausible maximum 
backsplash radii) the result does not vary considerably: 
$-10.97 \pm 0.92$ with an inter-quartile range of 1.78.

We next turn to the core population.  There may be 
some ambiguity in defining the core population due to 
infall interlopers at $R<R_{virial}$ whose three dimensional
radius from the cluster centre is $>R_{virial}$ 
(Diaferio et al.\ 2001; Rines et al.\ 2005).  The fraction
of such interlopers is expected to be low, however (perhaps 
10 to 20\%).  Therefore in order to get a handle on
the specific star formation rates in the cluster core,
we restrict ourselves to not only small radii from the
cluster centres ($R<0.5R_{virial}$), but also 
to those galaxies that are highly centrally peaked
(explicitly: $|\Delta cz| / \sigma_{cz}<0.3$).
This is reasonable given the predictions of the large
number of backsplash galaxies and particles that have visited 
this zone (Balogh et al.\ 2000; Moore et al.\ 2004). 
For these galaxies, the mean SSFR is found to be 
$log_{10} (SSFR) = -11.50 \pm 0.73$ with an inter-quartile 
range of 0.67.

\begin{figure}
\centerline{\psfig{file=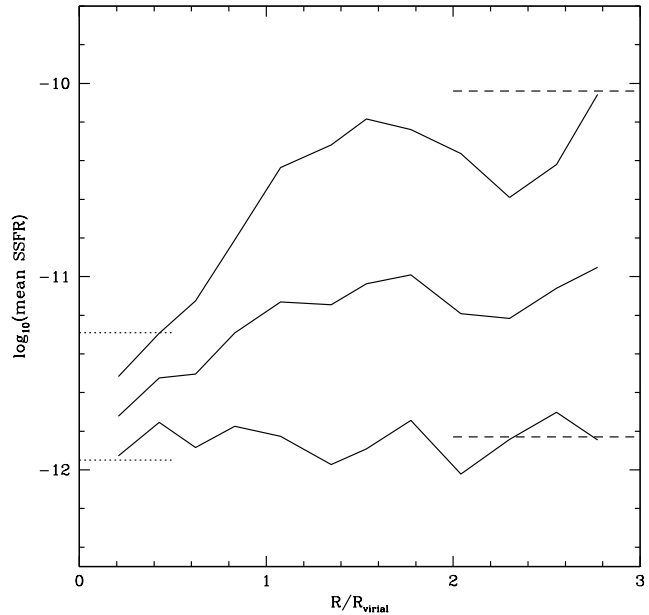,angle=0,width=3.5in}}
  \caption{Variation of the mean and inter-quartile SSFR range 
of galaxies in the range $0.3<|\Delta cz| / \sigma_{cz}<0.5$ 
with radius from the cluster centre (solid lines).  The 
inter-quartile range for first time infallers ($R>2R_{virial}$)
and the core population 
($R<0.5R_{virial}$ and $|\Delta cz| / \sigma_{cz}<0.3$) are displayed
for reference as the horizontal dashed and dotted lines respectively.
We can model the change in radius as a mixture of these infalling and 
core populations.
}
\label{fig:ssfr}
\end{figure}

In Fig.~\ref{fig:ssfr}, we demonstrate how the specific star
formation rate of the mixed population 
($0.3<|\Delta cz| / \sigma_{cz}<0.5$) varies with distance
from the cluster centre.  In line with other results (cf.\ G{\'o}mez
et al.\ 2003), we observe a 
decline of SSFR at the centre of the composite cluster.
Importantly, from Fig.~\ref{fig:ssfr},
it appears to be the case that the variation of the mixed
galaxy population with radius can be 
effectively thought of as a superposition of the
core and infalling population.

%
%

\subsection{Mixture Modelling}

%
%
\begin{figure}
\centerline{\psfig{file=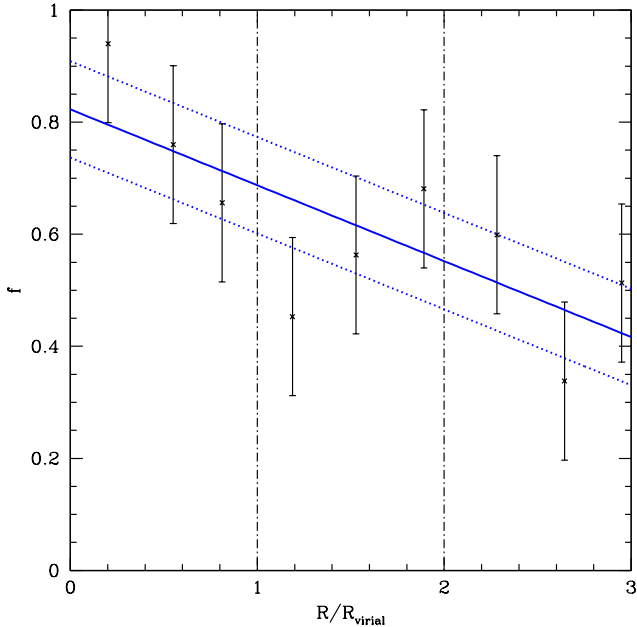,angle=0,width=3.5in}}
  \caption{Radial fraction of galaxies 
in the $0.3< |\Delta cz| / \sigma_{cz} <0.5$ slice
consistent with the distribution of
SSFR of the low velocity offset core galaxies 
(Fig.\ref{fig:ssfr}) that we interpret as the backsplash fraction.
Each point contains 50 galaxies and the errorbar is
the poisson error for this.
The solid line shows the linear line of best fit
($f = -0.135 R/R_{virial} +0.823$) to the data points
together with $1\sigma$ errors on the fit's intercept (dotted lines;
$\pm 0.086$).
The vertical dot-dash lines denote the expected mixed population
radial regime ($1<R/R_{virial}<2$) where these fractions
are expected to be of maximum utility (i.e.\ at $>> 2R/R_{virial}$
there are low, to zero backsplash galaxies expected).  
These fractions are  
consistent with previous estimates in other works.
}
\label{fig:fraction}
\end{figure}

We have thus far not attacked the central question
of the relative fraction of backsplash galaxies with
radius in the composite cluster.  
To do this, we will need to evaluate what relative proportions
of galaxies in the composite cluster have SSFRs that are
consistent with either the core population 
(i.e.\ consistent with having their star formation rates
truncated due to interactions with the hostile cluster core)
or the infalling
population as computed above (cf.\ Fig~\ref{fig:ssfr}).
This can be readily 
achieved using a mathematical mixture model.  
One of the most frequently used mixture models in galaxy cluster
analysis is Kaye's Mixture Model
(KMM; Ashman et al.\ 1994).  Although KMM is extensively
used in the literature to segregate substructure in clusters
(e.g.\ Pimbblet et al.\ 2010; Owers et al.\ 2009), it is
very much a general purpose tool that can be applied 
to the present situation.  

As Gaussian inputs to the KMM algorithm we supply the means
and standard deviations of the SSFR
for the core and infalling population that we found above.
Then for radial bins in the  
$0.3< |\Delta cz| / \sigma_{cz} <0.5$ slice, we implement KMM to 
partition the data and inform us of the relative fractions of 
galaxies that are consistent with either 
the low velocity offset, core region population or infalling 
population (as defined above).
The result of this analysis is presented in Fig.\ref{fig:fraction}.
Although there is scatter in the points, the trend
shows a downward trajectory with radius.  In the
mixed population radial regime ($1<R/R_{virial}<2$), we find
that the fraction of galaxies with SSFRs similar to the
low velocity offset cluster core region that we are interpreting
as backsplash galaxies varies
from $0.68 \pm 0.09$ (at $R_{virial}$) to $0.55 \pm 0.09$
(at $2R_{virial}$).  

We note that the relationship presented in 
Fig.\ref{fig:fraction} tells us only about the fraction of galaxies
that have SSFRs consistent with the low velocity core galaxies.
If we \emph{interpret} these as backsplash galaxies (i.e.\ by 
assuming they have visited the hostile cluster core and had
their rates truncated), we caution that the relation should
only be applied at radii where mixed backsplash and infall
galaxy populations are expected:
especially $1<R/R_{virial}<2$.  
Outside of this regime ($>> 2 R/R_{virial}$),
there are only low (or zero) backsplash galaxies expected 
(cf.\ Mamon et al.\ 2004) whereas below $R_{virial}$,
we are only detecting the core population that we
modelled the backsplash galaxies on in the first
instance.

We now ask whether our calculated backsplash values agree with 
previous works?  One of the few predictions on this value 
comes from Gill et al.\ (2005; their Fig.~8). 
That prediction (reproduced in Rines et al.\ 2005; their Fig.~8, top panel) 
show that the backsplash fraction is expected to be $\approx0.62$ at 
$1.4<R/r_{200}<2.8$ and $|\Delta cz| / \sigma_{200} \sim 0.4$.
This is in agreement with our best fit line (Fig.~\ref{fig:fraction}).

\section{Discussion}
\begin{figure}
\centerline{\psfig{file=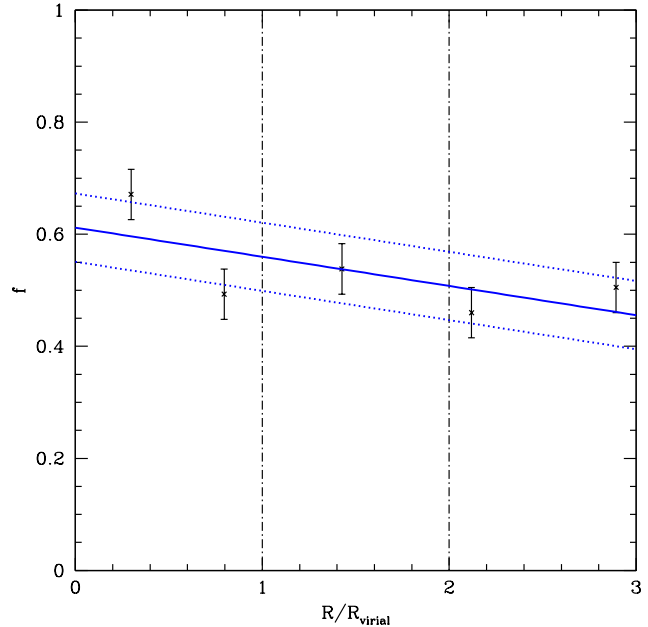,angle=0,width=3.5in}}
  \caption{As for Fig.~\ref{fig:fraction}, but with no
cut made on $|\Delta cz| / \sigma_{cz}$ and each point
containing 500 galaxies.  The observed trend 
($f = -0.052 R/R_{virial} +0.612$) is shallower
than that depicted in Fig.~\ref{fig:fraction}.
}
\label{fig:fraction2}
\end{figure}

To extend the utility of our results, we show in Fig.~\ref{fig:fraction2}
how the fraction varies as a function of radius for the
entire cluster population -- i.e.\ with no cut made on 
$|\Delta cz| / \sigma_{cz}$ other than they fall within 
the cluster membership caustic (Fig.~\ref{fig:stacked}).  
The form of the slope 
($f = -0.052 R/R_{virial} +0.612$) is much shallower than
for the $0.3<|\Delta cz| / \sigma_{cz}<0.5$ 
slice (Fig.~\ref{fig:fraction}). At $R_{virial}$, 
Fig.~\ref{fig:fraction2} suggests a fraction of
$0.59 \pm 0.06$ galaxies are backsplash, dropping
to $0.52 \pm 0.06$ at $2R_{virial}$.
We suspect the reason for this shallower relationship
is the increase in the infalling interloper fraction
due to making no cut in $|\Delta cz| / \sigma_{cz}$.
Indeed, Rines et al.\ (2005) demonstrate that up to 
60\% of emission line galaxies below $r_{200}$ 
($\sim R_{virial}$) are
probably interlopers that have a three dimensional 
radius from the cluster of $>R_{virial}$.  
Given that such emission line galaxies have significantly
higher values of $|\Delta cz| / \sigma_{cz}$ in clusters
(Pimbblet et al.\ 2006; Rines et al.\ 2005; 
see also Biviano \& Katgert 2004),
we regard it as very likely that this is the
prime cause of the change between Fig.~\ref{fig:fraction}
and Fig.~\ref{fig:fraction2}.

%
%

\begin{figure}
\centerline{\psfig{file=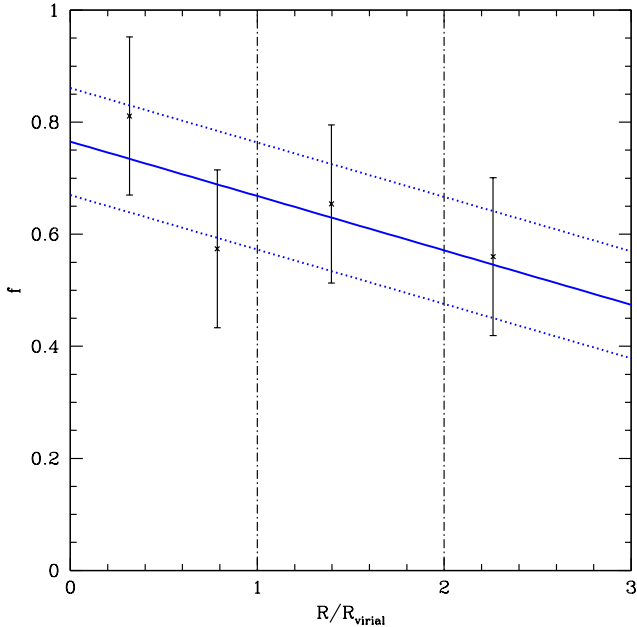,angle=0,width=3.5in}}
  \caption{As for Fig.~\ref{fig:fraction}, but for
the reduced, volume limited sample of 6 clusters
with no cut made on $|\Delta cz| / \sigma_{cz}$. 
Each point contains 50 galaxies.  
The fitted trend is of the form
$f = -0.097 R/R_{virial} +0.765$.
}
\label{fig:fraction3}
\end{figure}

From the outset, we caution that 
the sample of galaxies that we have used has 
not been volume limited.  Although we may have, in principle,
circumvented this by the use of specific star formation
rates (Fig.~\ref{fig:ssfr}), it remains the case that 
some environmental galaxy evolution mechanisms may
preferentially operate on lower mass galaxies 
at different radii and densities than 
more massive ones (e.g.\ Penny et al.\ 2010 and
references therein).  However, if anything, we suspect that
the star formation suppression that occurs in lower
mass galaxies may be more severe than in high mass
ones (e.g.\ Boselli et al.\ 2008), leading to a higher
fraction of backsplash galaxies in dwarf regimes at
comparable radii than for their more massive cousins.
To test this and 
create a volume-limited sample, we must necessarily
truncate the clusters used to the same portion of 
the luminosity function.  For SDSS spectroscopy
of the main galaxy sample, 
we note that the limiting magnitude for $\sim100$\% 
completeness is $r=17.77$ (Abazajian et al.\ 2009;
Strauss et al.\ 2002).  We use a
simple stellar population 
coupled with a solar metallicity and $z=5$ formation
time as input to Bruzual \& Charlot's (2003) code 
to predict how this magnitude limit varies as a function
of redshift.  Between our highest redshift cluster (A0655)
and lowest (A0119), the $r=17.77$ limit corresponds to
a change in absolute magnitude of $\Delta M_R\sim2.5$.
Such a large change would mean that our lowest redshift 
clusters would contribute only very small numbers of galaxies
($\sim$1's to 10's of galaxies) to the composite, and overall we
would be reduced to a small ($< 100$ galaxies) sample size.  
To alleviate this issue, 
we consider clusters in a narrower redshift slice: 
from A1620 ($z=0.084$) to A1767 ($z=0.070$).  This
reduced sample totals 6 
clusters and spans a change of absolute
magnitude of $\Delta M_R\sim 0.4$.  For A1620 (the 
highest redshift cluster in the reduced range), the
SDSS apparent magnitude completeness limit
corresponds to an $\approx M^{\star}+2$ galaxy.  
We now truncate
the galaxies in the other 5 clusters in the narrower
redshift regime to the same limit and reproduce the
above analysis on the reduced sample.  
Fig.~\ref{fig:fraction3} displays these results.  
At $R_{virial}$, the backsplash fraction from
the fitted line is $0.67 \pm 0.10$, decreasing
to $0.57 \pm 0.10$ at $2R_{virial}$.
We therefore regard these volume limited results 
as compatible (i.e.\ within $1\sigma$) 
with the previous results 
(Fig.~\ref{fig:fraction2}).

%
%

Lastly, we note here that changing the exact choice of the
cluster centre to (e.g.) the brightest cluster member
has no significant effect on the fits given in Figs.~4
through 6.

\section{Summary}

The main findings of this work are summarized below.

(1) We have detected a mixed population of infalling
and backsplash galaxies in isolated (in both redshift space
and the plane of the sky) galaxy clusters that are free
from recent cluster-cluster merger activity.

(2) We have 
modelled the backsplash population on 
low velocity offset core population members under the
assumption that backsplash galaxies will have 
had their SSFRs reduced from
interaction with the hostile cluster core.  This
is reasonable given the predictions of the sheer
numbers of backsplash galaxies that have visited
this zone (cf.\ Moore et al.\ 2004).
By using a mixture model,
we suggest that this population
follows the relation 
$f = -0.052R/R_{virial} + 0.612 \pm 0.06$
for all cluster members.

We explicitly note that 
this equation should only be applied where there is
a mixed population expected -- i.e.\ not in the very high
density core, and not at the outskirts of the cluster
where backsplash galaxies are unexpected to reside.
Ideal radii for application is over the range $\sim R_{virial}$
to $\sim 2 R_{virial}$.  We caution that this relation
makes no account of infall interlopers.

(3) Our results are in broad agreement with previous
observational 
works in this area (e.g.\ Pimbblet et al.\ 2006; Rines et al.\ 2005) 
who do not use such idealized isolated samples.  

Taken together, our results support the viewpoint that 
mechanisms tied to local galaxy density should be more important
drivers of star formation than distance to galaxy clusters.
This is in agreement with, and adds weight to,
our earlier results (Pimbblet et al.\ 2006) 
and is opposite to (e.g.) Whitmore, Gilmore \& Jones (1993).

In the near future, it will be interesting to closely 
examine the results of the Galaxy and Mass Assembly 
survey (GAMA; Driver et al.\ 2010) to see what
is happening in the more dwarfier regimes, and the on-going
investigations of Owers et al.\ (2009) who have obtained very
high spectroscopic completeness in a batch of clusters.

\section*{Acknowledgements}
KAP thanks Michael Brown, John Stott, Isaac Roseboom 
\& Matt Owers for useful conversations during the course of this work 
and also Ruth Pimbblet for inspiring it. 

I would like to express my gratitude to the referee, Alexander Knebe,
for rapid, useful comments and constructive advice on the earlier 
version of this manuscript.

This work has been generously 
supported through Monash University
grant number 3934275.

Funding for the SDSS and SDSS-II has been provided by the
Alfred P. Sloan Foundation, the Participating Institutions,
the National Science Foundation, the U.S. Department of Energy,
the National Aeronautics and Space Administration,
the Japanese Monbukagakusho, the Max Planck Society, and
the Higher Education Funding Council for England.

The SDSS is managed by the Astrophysical Research Consortium for
the Participating Institutions. The Participating Institutions are
the American Museum of Natural History, Astrophysical Institute Potsdam,
University of Basel, Cambridge University, Case Western Reserve University,
University of Chicago, Drexel University, Fermilab,
the Institute for Advanced Study, the Japan Participation Group,
Johns Hopkins University, the Joint Institute for Nuclear Astrophysics,
the Kavli Institute for Particle Astrophysics and Cosmology,
the Korean Scientist Group, the Chinese Academy of Sciences (LAMOST),
Los Alamos National Laboratory, the Max-Planck-Institute for Astronomy (MPIA),
the Max-Planck-Institute for Astrophysics (MPA),
New Mexico State University, Ohio State University, University of Pittsburgh,
University of Portsmouth, Princeton University,
the United States Naval Observatory, and the University of Washington.

\end{document}